\documentclass[conference]{IEEEtran}
\IEEEoverridecommandlockouts
\usepackage{cite}
\usepackage{amsmath,amssymb,amsfonts}
\usepackage{graphicx}
\usepackage{textcomp}
\usepackage{xcolor}

\usepackage{mathtools}
\usepackage{subcaption} 
\usepackage{multirow}
\usepackage{tikz}
\usepackage{pgfplots}
\usepackage{url}
\usepackage{balance}
\pgfplotsset{compat=1.14}

\usepackage[normalem]{ulem}
\usepackage{balance}
\usepackage{enumitem}
\usepackage{soul}
\usepackage{tikz}

\usepackage{algpseudocode}
\usepackage{algorithm}

\usepackage{caption} 
\usepackage{textcomp}
\newtheorem{analysis-rule}{Rule}[subsection]
\newtheorem{analysis-sub-rule}{Rule}[analysis-rule]

\usepackage{listings}
\usepackage{color}
\usepackage{cite}
\usepackage{amssymb}

\DeclareMathAlphabet{\mathpzc}{OT1}{pzc}{m}{it}

\definecolor{dkgreen}{rgb}{0,0.6,0}
\definecolor{gray}{rgb}{0.5,0.5,0.5}
\definecolor{mauve}{rgb}{0.58,0,0.82}

\newcommand{\oursys}{\textsc{PPrior}}

\newcommand\redsout{\bgroup\markoverwith{\textcolor{red}{\rule[0.5ex]{2pt}{0.4pt}}}\ULon}

\definecolor{fgreen}{rgb}{0.0, 0.5, 0.0}

\newcommand{\stitle}[1]{\vspace{1ex}\noindent\textup{\textbf{#1}}}

\newcommand{\CASE}[1]{\STATE \textbf{case} #1\textbf{:} \begin{ALC@g}}
\newcommand{\ENDCASE}{\end{ALC@g}}

\newcommand{\DEFAULT}{\STATE \textbf{default:} \begin{ALC@g}}
\newcommand{\ENDDEFAULT}{\end{ALC@g}}
\newcommand{\DEFAULTLINE}[1]{\STATE \textbf{default:} }
\makeatletter
\newcommand*\bigcdot{\mathpalette\bigcdot@{.5}}
\newcommand*\bigcdot@[2]{\mathbin{\vcenter{\hbox{\scalebox{#2}{$\m@th#1\bullet$}}}}}
\makeatother

\newcommand{\myspecial}[1] {\texttt{#1}}
\newcommand{\ourmodel}{$\mathsf{PPrior}$}
\usepackage{comment}

\usepackage{multirow}
\usepackage{array}

\def\BibTeX{{\rm B\kern-.05em{\sc i\kern-.025em b}\kern-.08em
    T\kern-.1667em\lower.7ex\hbox{E}\kern-.125emX}}
\begin{document}

\IEEEoverridecommandlockouts
  \IEEEpubid{\makebox[\columnwidth]{
  978-1-6654-8045-1/22/\$31.00 \copyright 2022 IEEE} 
 \hspace{\columnsep}\makebox[\columnwidth]{ }}

\title{Proactive Prioritization of App Issues via Contrastive Learning}

\author{\IEEEauthorblockN{Moghis Fereidouni\IEEEauthorrefmark{1}, Adib Mosharrof\IEEEauthorrefmark{1}, Umar Farooq\IEEEauthorrefmark{2}, A.B. Siddique\IEEEauthorrefmark{1}}
\IEEEauthorblockA{\IEEEauthorrefmark{1}
    Department of Computer Science,%
    University of Kentucky, 
    \IEEEauthorrefmark{2} Independent Researcher \\
 Email: 
mfe261@uky.edu, amo304@g.uky.edu, ufarooq.cs@gmail.com, siddique@cs.uky.edu
}}

\maketitle
%\thispagestyle{plain}
%\pagestyle{plain}
%\begingroup\renewcommand\thefootnote{\textasteriskcentered}
%\footnotetext{Equal contribution.}
%\endgroup
\begin{abstract}
Mobile app stores produce a tremendous amount of data in the form of user reviews, which is a huge source of user requirements and sentiments; such reviews allow app developers to proactively address issues in their apps.
However, only a small number of reviews capture common issues and sentiments which creates a need for automatically identifying prominent reviews.
Unfortunately, most existing work in text ranking and popularity prediction focuses on social contexts where other signals are available, which renders such works ineffective in the context of app reviews.
In this work, we propose a new framework, {\ourmodel}, that enables proactive prioritization of app issues through identifying prominent reviews (ones predicted to receive a large number of votes in a given time window).
Predicting highly-voted reviews is challenging given that, unlike social posts, social network features of users are not available.
Moreover, there is an issue of class imbalance, since a large number of user reviews receive little to no votes.
{\ourmodel} employs a pre-trained T5 model and works in three phases. 
Phase one adapts the pre-trained T5 model to the user reviews data in a self-supervised fashion. In phase two, we leverage contrastive training to learn a generic and task-independent representation of user reviews. 
Phase three uses radius neighbors classifier to make the final predictions. 
This phase also uses FAISS index for scalability and efficient search. 
To conduct extensive experiments, we acquired a large dataset of over 2.1 million user reviews from Google Play.
Our experimental results demonstrate the effectiveness of the proposed framework when compared against several state-of-the-art approaches.
Moreover, the accuracy of {\ourmodel} in predicting prominent reviews is comparable to that of experienced app developers.

\end{abstract}

% \begin{IEEEkeywords}
% App reviews, response generation, neural machine translation.
% \end{IEEEkeywords}


\section{Introduction}
\label{sec:introduction}

\begin{figure}[t!]
\centering
\includegraphics[width=0.95\linewidth]{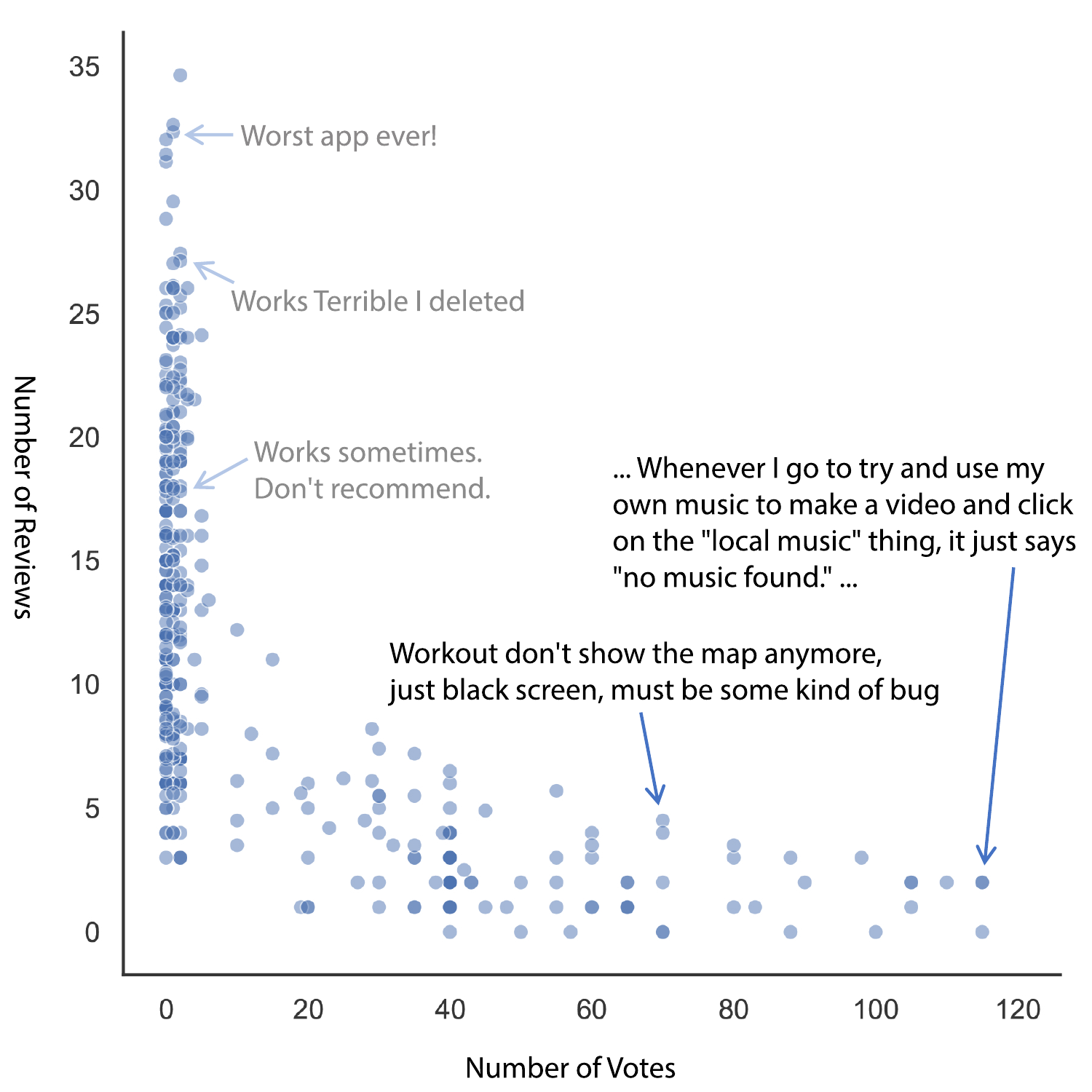}
\caption{Following the power law distribution, a small percentage of user reviews receive higher votes, and most reviews get little to no attention.
Highly-voted reviews help app developers prioritize features and fix bugs because they typically highlight precise issues.
}
\vspace{-0.8cm}
\label{fig:intro}
\end{figure}

The popularity of mobile apps has grown exponentially over the last few years, and according to a recent study, users spend more than four hours a day using mobile apps in the United States~\cite{statista:time-spent}.
Such user activity results in lucrative user feedback, which is usually shared on the app stores such as Google Play and Apple App store.
In addition to posting reviews, these distribution channels allow users to like or dislike other users' reviews.
The user reviews include a range of information, including issues with apps, suggestions for refinements of apps, and feature requests, among others.
It is important for app developers to acknowledge user reviews (i.e., write review responses) and address users' concerns to stay competitive and retain users.
While some previous work has focused on automatic review response generation~\cite{rrgen,aarsyth}, identifying critical user feedback from a massive amount of reviews remains a challenge.

There are several characteristics of user reviews that make them challenging to analyze.
First, manually processing user reviews is virtually impossible due to their volume, velocity, and voracity.
Second, user reviews are typically written using mobile phones and often contain noisy words (e.g., misspelled, repetitive, non-English). 
Existing research in this direction has focused on pre-processing reviews~\cite{keyword-approach}, filtering reviews~\cite{chen2014ar}, and topic classification~\cite{di2016would}.
Nonetheless, it is difficult to identify critical app issues automatically from reviews, since problems and concerns frequently shift over time and change across apps.

\begin{figure*}[t!]
\centering
\includegraphics[width=0.95\linewidth]{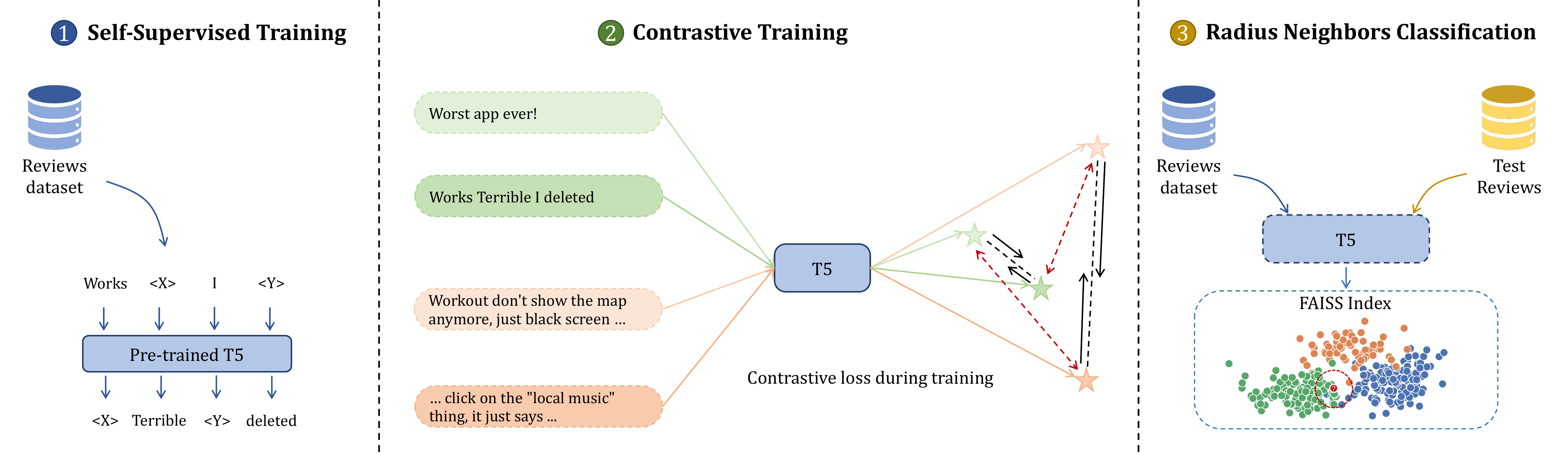}
\caption{Overview of {\ourmodel}. Contrastive training in phase two is a critical component of the proposed framework.}
\vspace{-14pt}
\label{fig:intro_method}
\end{figure*}

%in a certain amount of time
In this work, we propose a novel framework, {\ourmodel}, for \textbf{P}roactive \textbf{Prior}itization of App issues by automatically predicting the number of votes (i.e., the number of people who found that review useful) a  given review will receive, allowing app developers to \emph{proactively} prioritize those app issues.
{\ourmodel} builds upon the following insights.
First, many users of an app may face similar issues and share their concerns. 
% Second, fixing a few software issues could possibly fix a  majority of the bugs and problems~\cite{microsoft-8020}.
Second, fixing a few software issues faced by early users could possibly fix the majority of bugs and problems for all users~\cite{microsoft-8020}.
Third, it is in the interest of app developers to prioritize features or bug fixes that most users would like to see, following the maximum return criterion.
A preliminary study on a subset of a large-scale dataset of user reviews from Google Play also confirms these insights, where a small number of reviews receive a large number of votes, usually expressing very specific and precise concerns, which many users experience. 
To highlight this phenomenon, Figure~\ref{fig:intro} shows a subset of user reviews. 
We argue that developers should prioritize resolving this small subset of issues given their impact on a large number of users.
Figure~\ref{fig:intro} also shows a few sample reviews that received little to no votes. 
For example, among the reviews with negative sentiment, ``Worst app ever!'', the developer can not make much of such reviews since they do not highlight any particular issues with the app.
This work aims to accurately predict the number of votes a negative user review will receive in a certain amount of time to \emph{proactively} prioritize those issues. 
That is, proactively resolving app issues before they lead to frustrating a large number of users and result in firefighting situations.

We formulate the task of predicting the number of votes as a classification task, which seemingly resembles predicting the popularity of social media posts, e.g., the number of likes or retweets for a Facebook post or a Twitter tweet, respectively. 
It turns out, however, that those tasks are inherently different since the existing research~\cite{khoerunnisa2022prediction,saeed2022framework,daga2020prediction} in popularity prediction of social media posts relies heavily on the user’s social network features that are not available on app stores where the number of votes is purely determined based on the content of the user review and whether other users also share similar concerns or not, which makes this task practically more relevant, interesting, and challenging.

Our proposed framework, {\ourmodel}, leverages the power of pre-trained language models, sentence transformers~\cite{reimers2019sentence}, and employs contrastive learning. 
Figure~\ref{fig:intro_method} presents an overview of the framework that works in three phases and uses a pre-trained T5~\cite{raffel2020exploring} as the backbone model. 
Language models, such as T5, are trained on massive amounts of text data in a self-supervised fashion and have millions of parameters.
Therefore, they are capable of generating rich and accurate representations of any textual data.
Phase one employs the pre-trained T5 and further performs self-supervised training.
Phase two adapts a contrastive training objective. Contrastive learning enables task-independent training and forces the model to learn a generic embedding space where similar user reviews are located close together and dissimilar reviews are spread out. 
The user reviews with zero or a small number of votes generally do not highlight any specific concern, thus ending up being close to each other in the high-dimensional embedding space. 
Moreover, the generic embedding space also takes care of the distributional shifts in the reviews.
For efficient similarity search, we store the training samples using Facebook AI Similarity Search (FAISS)~\cite{johnson2019billion} and use radius neighbors classifier -- a better choice as compared to classical KNN when data is not uniformly sampled~\cite{zhu2015gravitational} -- to predict the number of votes for a given user review in phase three.

To evaluate the effectiveness of {\ourmodel}, we first crawled a dataset that consists of over 2.1 million negative user reviews from about 10,000 Google Play apps~\cite{googleplay}. 
We also recorded the number of votes each review received in a month after being posted because the number of votes a user review receives is also dependent on the length of time that has passed since the review was posted.
Existing user reviews datasets~\cite{rrgen,aarsyth} do not cover a large number of apps. 
Moreover, these datasets do not record the number of votes for reviews after a fixed amount of time after they were posted. 
We conduct extensive experimental evaluations using the large dataset and compare against several state-of-the-art transformer architectures~\cite{devlin2018bert,brown2020language,song2020mpnet,raffel2020exploring} that leverage several strategies to overcome the class imbalance issue~\cite{lin2017focal,sudre2017generalised}. 
We also compare against state-of-the-art approaches for predicting the social media post's popularity~\cite{saeed2022framework,daga2020prediction}.
Our results show that {\ourmodel} outperforms all the competing approaches by at least 27.97\% and 24.50\% on Matthew's correlation coefficient (MCC)~\cite{chicco2020advantages} score for binary and multi-class classification tasks, respectively. 
Moreover, our human study shows that the performance of {\ourmodel} is comparable to experienced software developers in identifying critical app issues.
The source code and dataset are available publicly\footnote{\url{https://github.com/MultifacetedNLP/PPrior}}.

In summary, the contributions of this paper are as follows:
\begin{itemize}[leftmargin=6mm]

\item
We release a big dataset containing over 2.1 million negative user reviews from Google Play for around 10,000 apps along with the number of votes each review received in a month.

\item
We introduce a novel framework for predicting the number of votes for a specific user review in order to enable proactive prioritization of app issues.

\item
We use our large dataset to conduct extensive experiments and compare {\ourmodel} with state-of-the-art methods. Automatic metrics and real user studies confirm {\ourmodel}'s competitiveness with significant improvements.
\end{itemize}

The remainder of the paper is organized as follows. 
We present the preliminaries in Section~\ref{sec:background}, {\oursys} in Section~\ref{sec:model}, followed by our experimental setup in Section~\ref{sec:experiments} and evaluation results in Section~\ref{sec:evals}. Finally, we discuss the related work in Section~\ref{sec:related} and conclude the paper in Section~\ref{sec:conclusion}.
\section{Preliminaries}
\label{sec:background}
Our proposed framework, {\ourmodel}, builds upon several mature components from natural language processing (NLP) literature. We provide a brief overview of those in the following.

\stitle{Unsupervised Representation Learning and Pre-trained Language Models.}
Unsupervised (or self-supervised) latent representation learning~\cite{siddique2020unsupervised} and pre-trained language models have contributed greatly to recent NLP success
~\cite{raffel2019exploring, devlin2018bert, song2020mpnet,radford2019language}, including facilitating zero-shot learning~\cite{siddique2021unsupervised,siddique2022personalizing}.
The unsupervised learning technique enabled the development of more robust NLP systems due to the abundance of textual data. 
We need to optimize the proposed framework using unsupervised representation learning techniques to handle distributional shifts~\cite{siddique2021linguistically,siddique2021generalized} in a robust manner so that it can handle new unseen app issues accurately.
The language models (e.g., GPT-2~\cite{radford2019language}, BERT~\cite{devlin2018bert}, T5~\cite{raffel2020exploring}) are trained using unsupervised techniques by utilizing vast amounts of text data. 
These models can capture both general semantic and syntactic information effectively, due to the size of the model and training dataset.
In this work, we employ T5, a pre-trained language model, as the backbone model in our framework.
We also leverage a variation of  Sentence-BERT~\cite{reimers-2019-sentence-bert} architecture to learn sentence-level representations.  
Instead of BERT, we use a more powerful pre-trained language model T5 for learning sentence embeddings. The T5 model follows transformers~\cite{vaswani2017attention} architecture, contains both encoder and decoder stacks, and has produced state-of-the-art results on several NLP benchmarks, including SuperGLUE.% and SQuAD~\cite{rajpurkar2016squad}.
%siddique2021generalized

\stitle{T5 Model.} 
Inspired by the idea of casting every NLP task as question answering~\cite{mccann2018natural}, T5 treats every task as a text-to-text problem. 
With a text-to-text framework, virtually every task can be addressed with the same model, objective, training process, and decoding process. 
T5 Adapts the original transformers architecture for multi-task learning, as it has both encoder and decoder stack as opposed to BERT (i.e., only encoder stack is more suitable for natural language understanding tasks) and GPT (i.e., only decoder stack is more appropriate for natural language generation tasks).
The transformers architecture is primarily based on self-attention block~\cite{cheng2016long} -- an attention variant where each element of the sequence is replaced by a weighted average of the remaining elements in the sequence~\cite{graves2013generating,bahdanau2014neural}.
T5 encodes text using SentencePiece~\cite{kudo2018sentencepiece} as WordPiece tokens~\cite{sennrich2015neural} and adapts the denoising and corrupting span objective that is inspired by masked language modeling and word dropout technique~\cite{bowman2015generating} in BERT, to train the model to predict missing spans (or corrupted spans) of text in the input. Figure~\ref{fig:intro_method} illustrates an example of the input and output of the T5 model. The example text, ``Works Terrible I deleted'', the input becomes, ``Works $<$X$>$ I $<$Y$>$'', after
replacing the dropped-out spans with unique single sentinel tokens ``$<$X$>$'' and ``$<$Y$>$''.
Then, the target is to predict only sentinel tokens and all other input tokens are replaced by the sentinel token ``$<$X$>$'' and ``$<$Y$>$''. 
In our example, the output becomes, ``$<$X$>$ Terrible $<$Y$>$ deleted''. 
Predicting only sentinel tokens is computationally cheap as compared to predicting everything in the target.
The T5 model is trained  on the colossal clean
crawled corpus (C4) dataset (about 750 GB of data). The C4 is a cleaner version of text extracted from the web in April 2019.
In this work, we employ T5 as a backbone model in our framework and fine-tune it in three phases to capture generic patterns that are robustly transferable to the test dataset. 
Our choice to use T5 is mainly guided by its multi-task learning capabilities and robustness to effectively learn in the distributional shifts scenarios.
We perform self-supervised training to adapt the model to the reviews dataset in phase one, further train it for the contrastive loss in phase two, and finally leverage the radius neighbors classifier to make final predictions about reviews. %in phase three.

\stitle{FAISS and KNN.} 
The Faiss library facilitates efficient similarity searches and clustering of dense vectors and supports a range of comparison operations, including L2 distance, dot product, and cosine similarity.
By adding an indexing structure on top of raw vectors using scalable approaches such as hierarchical navigable small worlds~(HNSW)~\cite{malkov2018efficient} and navigating spreading-out graphs~(NSG)~\cite{FuNSG17}, FAISS enables effective searching of billions of vectors.
In terms of implementation, it is implemented in C++ for the most part, with only BLAS as its dependency.
Further speeding up the inference can be achieved with GPU (both single and multi-GPU) indexes, supported via CUDA.
Due to its Python interface, it is compatible with all deep learning frameworks.
In this work, we employ FAISS to index our training dataset in phase three of our proposed framework {\ourmodel}.
Then, we use the radius neighbor classifier to make predictions.
The radius neighbor classifier makes predictions by taking the most common label among the neighbors in the given radius. 
Since there is a class imbalance in our reviews dataset, this is better suited compared to the classical KNN algorithm. Moreover, phase three does not need any training and thus any other algorithm can be plugged-in depending on the distribution of the data at inference time. In our experiments, we also use weighted KNN as an alternative.

\begin{figure*}[t!]
\centering
\includegraphics[width=\linewidth]{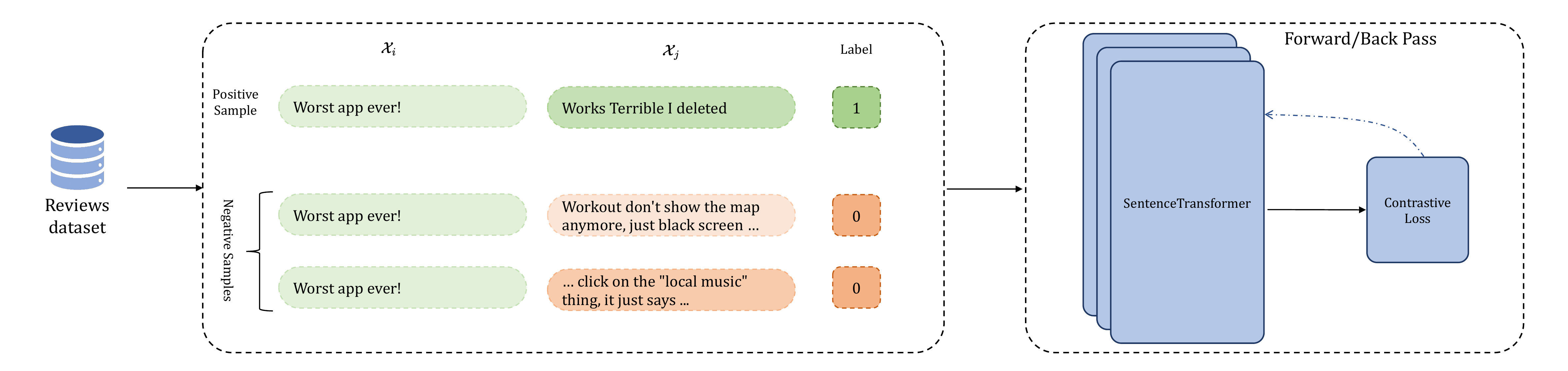}
\caption{Phase two of {\ourmodel} framework. The contrastive training enables learning generic representations of the user reviews where similar reviews are forced to stay close to one another. These robust and generic representations empower simple methods like KNN to identify critical user concerns effortlessly.}
\vspace{-3pt}
\label{fig:cmodel}
\end{figure*}

\section{Proactive Prioritization Framework: {\ourmodel}}
\label{sec:model}
In this work, we introduce {\ourmodel}, a novel framework for proactively prioritizing app issues by automatically predicting how many votes a given review will receive, which enables app developers to proactively prioritize those app issues.
The number of votes a user review receives is a direct observation of the number of people who found that review useful. 
{\ourmodel} employs the pre-trained T5 model as a backbone model and works in three phases. Figure~\ref{fig:intro_method} presents an overview of the framework.
Phase one loads the pre-trained T5 and further performs self-supervised training on the reviews dataset using the original training objective of denoising and corrupting span objective.
As T5 is pre-trained on the C4 dataset, which is fundamentally different from the app reviews data, this step overcomes the data distribution shift issue.
Phase two adapts a contrastive training objective that promotes task-independent and generic representation learning.
Since user reviews with zero or few votes dominate the dataset, there is an issue of class imbalance.
The training in phase two also addresses this challenge by generating negative samples and pairing samples with a high number of votes to those with a small number of votes (i.e., the minority class is exposed more often to the model). Thus, the effect of the class imbalance is minimized.
Moreover, the generic embeddings space also takes care of the issue of the frequent issue shifts in user reviews.
Phase three of the framework does not require any training and uses the learned representations from phase two to make inference. 
This phase uses FAISS to store large-scale user reviews training data in a scalable index and performs efficient similarity search to make final predictions. This phase uses a radius neighbors classifier that has been shown to perform better when data is non-uniformly sampled. 
Since this phase does not require any training, the prediction algorithm can be replaced on the fly depending on the distribution of the test set. In the following, we provide further details about each phase of the {\ourmodel}.

\subsection{Phase One: Self-Supervised Training}
We exploit the power of the pre-trained language models by initializing phase one of our framework with the pre-trained T5 model. 
Our implementation uses \myspecial{t5-base} that has 220 million parameters. 
It has been pre-trained on C4 dataset (750 GB of English textual data) for  524,288 steps and the vocabulary size is 32,000 wordpieces. 
Both the
encoder and decoder contain 12 blocks, where each block contains self-attention (with 12 attention
heads), encoder/decoder attention, and a feed-forward layer. The embedding layer has 768 dimensions. Moreover, it also uses dropout with a probability of 0.1 for regularization.

Phase one kicks off the training of the {\ourmodel} by loading the pre-trained \myspecial{t5-base} and performing self-supervised training using the user reviews dataset to adapt the model to the distribution of the user reviews. 
This phase does not change the original training objective (of T5 model), which corrupts the input spans randomly and tries to predict the masked-out spans. We randomly sample and drop out 15\% of the tokens. The consecutive dropped-out token spans are replaced by unique special tokens, called  sentinel IDs. We use cross-entropy loss and maximum likelihood using teacher forcing~\cite{williams1989learning} with AdaFactor optimizer~\cite{shazeer2018adafactor}. 
The greedy decoding is used that chooses the tokens with the highest probability at every timestep.
The self-supervised training phase is also presented in Figure~\ref{fig:intro_method}~(left part).
The output of phase one is the trained
T5 model that has been adapted to the user reviews.

\subsection{Phase Two: Contrastive Training}
This phase initializes with the T5 that was adapted to the user reviews data (i.e., the output of phase one) to take care of the distributional shift between web data (T5 is pre-trained on web data) and user reviews data that may contain semi-technical language. 
We further fine-tune the T5 model with a contrastive objective.
Figure~\ref{fig:cmodel} presents an overview of the contrastive training.
For training {\ourmodel} in phase two, we leverage negative sampling to capture the correspondence between user reviews. 
That is, shift the high-dimensional representations of the user reviews that receive a similar number of votes closer to one another and push away the representations of user reviews with a huge gap in the number of votes they receive.
We achieve this by randomly choosing $K$ negative samples for each positive pair. 
The positive pair represents two training samples that receive a similar number of votes, whereas the negative pair contains two samples with a very different number of votes. Figure~\ref{fig:cmodel} shows how training data is prepared for this phase. 
We leverage contrastive loss to guide the training in phase two. 
Our training approach enables generic transferable associations between
user reviews that are capable of handling new unseen emerging issues in the apps. 
Moreover, negative sampling manages to expose the minority class samples (e.g., very few user reviews receive a high number of votes) to the model as much as majority class instances.
In the following, we provide further
details about the training process.

\begin{figure}[t!]
\centering
\includegraphics[width=\linewidth]{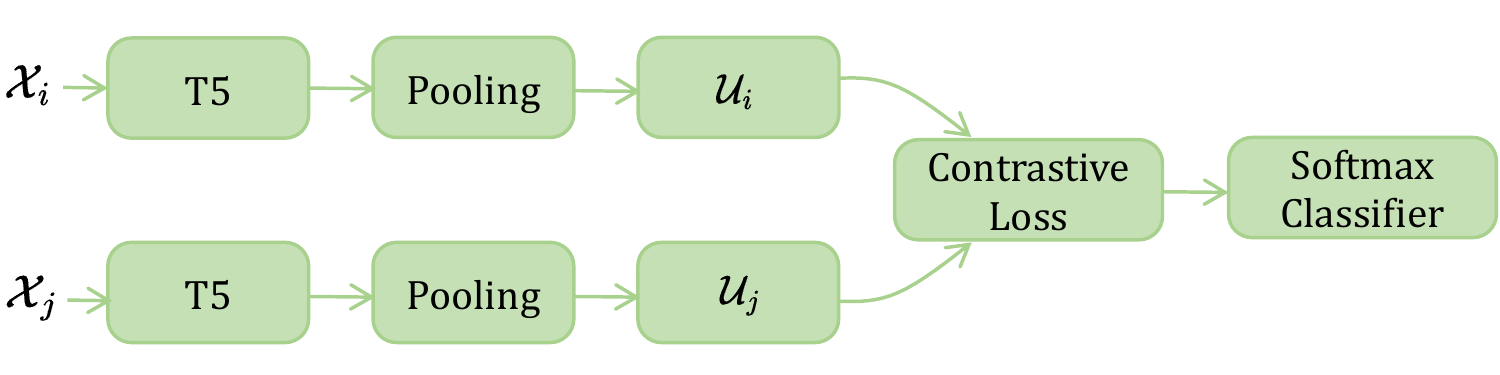}
\caption{Our modified sentence transformer uses T5 model.}
\label{fig:sentence-transformer}
\vspace{-8pt}
\end{figure}

\stitle{Sentence Transformer.}
The Sentence Transformer, also known as Sentence-BERT, was designed to solve the computational overhead issues involved in sentence pair regression task, encountered while using BERT\cite{devlin2018bert} and RoBERTa\cite{liu2019roberta} like architectures. 
The Sentence Transformer addressed this challenge by using siamese and triplet loss~\cite{schroff2015facenet}. 
In this work, we employ Sentence Transformer jointly with T5 for our contrastive learning-based training. 
We modified Sentence-BERT architecture by replacing the BERT model with our T5 model to calculate sentence embeddings. Figure~\ref{fig:sentence-transformer} represents the modified Sentence-BERT architecture with integrated T5.

\stitle{Negative Sampling.}
Assume that $X = \{\mathcal{X}_1, \mathcal{X}_2, \cdots, \mathcal{X}_n\}$ represents the user reviews and $y_i \in \mathbb{R}$ represents the number of votes $i$th review $\mathcal{X}_i$ received.
Given two training samples $\mathcal{X}_i$ and $\mathcal{X}_j$, we assign the label 1 if $|y_i - y_j| < \lambda$ (i.e., positive pair). 
For each positive pair $\mathcal{X}_i$ and $\mathcal{X}_j$, we replace $\mathcal{X}_i$ (or $\mathcal{X}_j$, one at a time) with an instance $\mathcal{X}_q$ such that $|y_i - y_q| \ge \lambda$ and assign it label 0 (i.e., negative pair). For each positive pair, we generate $K$ negative pairs. 
That is, for the given two samples, if the  absolute difference of the number of votes they received is smaller than $\lambda$, we assign labels of 1, and 0 otherwise. 
In our experiments, we use negative sampling of 1:4 and $\lambda$ is set to 100 and 4 for binary classification and multi-class classification tasks, respectively, with added constraints:
i) instances in a negative pair should not come from the same class, and ii) positive pair contains both instances from the same class.

\stitle{Input Embeddings.}
Let $\mathcal{F}_{\theta}$ denote our modified sentence transformer model and $(\mathcal{X}_i, \mathcal{X}_j)$ be a training example. We acquire the representations $\mathcal{U}_i$ and $\mathcal{U}_j$ for $\mathcal{X}_i$ and $\mathcal{X}_j$, respectively, using: 

\vspace{-10pt}
\begin{equation}
   \mathcal{U}_i = \mathcal{F}_{\theta}(\mathcal{X}_i)
\end{equation}

\begin{equation}
   \mathcal{U}_j = \mathcal{F}_{\theta}(\mathcal{X}_j)
\end{equation}
Where $\theta$ represents the model parameters.

\vspace{10pt}
\stitle{Objective Function.}
We use a contrastive objective function to fine-tune sentence-level embeddings. 
For the given positive training example $(\mathcal{X}_i, \mathcal{X}_j)$ with a label of 1 and negative training example $(\mathcal{X}_i, \mathcal{X}_q)$ with label 0, the constrastive loss forces the positive samples $\mathcal{X}_i$ and $\mathcal{X}_j$ closer together in the embedding space, while pushes away the negative samples $\mathcal{X}_i$ and $\mathcal{X}_q$.
After we acquire the reviews embeddings using the sentence transformer for the positive training example $(\mathcal{X}_i, \mathcal{X}_j)$ as $\mathcal{U}_i$ and $\mathcal{U}_j$, and negative training example $(\mathcal{X}_i, \mathcal{X}_q)$ as $\mathcal{U}_i$ and $\mathcal{U}_q$, the training loss can be defined as:

\begin{equation}
    \mathcal{L}_{i,j} = -\log\frac{\exp\left ( \mathcal{U}_i \bigcdot \mathcal{U}_j / \tau\right )}
    {\sum\limits_{q \in Q }\exp\left ( \mathcal{U}_i \bigcdot \mathcal{{U}}_q / \tau  \right )}
\end{equation}
where the $\bigcdot$ is the similarity function (e.g., dot product), $\tau\in\mathcal{R}^+$ is a scalar parameter for temperature, and $Q$ is the set of negative pairs.

\stitle{Generalization to New Unseen Reviews.}
The proposed contrastive objective design learns generic representations of the user reviews that focus on learning the association between reviews in such a way that review pairs with high affinity are pulled together, while those not bearing much similarity are pushed away in the high dimensional manifold.
By doing that, the model's objective is refined towards learning generic features, rather than learning the dataset-specific distribution and their corresponding characteristics. 
Moreover, the text-to-text framework of T5 and the generic nature contrastive learning enables multi-task learning. 
The trained T5 model in this phase can be used for a range of tasks such as binary classification, multi-class classification or regression without the need to re-train any component.

\subsection{Phase Three: Radius Neighbor Classification}
The phase three of the proposed framework is employed only for inference and does not require any training or fine-tuning.
Since the training dataset is huge in size, we use FAISS to index the user reviews for scalable and efficient retrieval of similar reviews that are used to make predictions. 
Specifically, we pass all the training reviews through the trained T5 (i.e., the output of phase two) and use the output of the last hidden layer as the embedding vector for the given user review.
An overview of FAISS has been provided in Section~\ref{sec:background} and interested readers are referred to~\cite{johnson2019billion} for details.
Phase three is presented in Figure~\ref{fig:intro_method} (right part). 
To make predictions, we use the radius neighbor classification (RNC) algorithm that uses neighbors in the given radius and selects the most common label as its predictions.
In our experimental evaluation, we conduct experiments for binary classification as well as multi-class classification tasks. Moreover, we also experiment with weighted KNN as a replacement for the radius neighbor classifier. We use a radius of 2 and choose the number of nearest neighbors to be 101.
Since this phase is only employed for inference, the indexing mechanism and classification algorithm can be changed effortlessly depending on the speed needs and data distribution.
 
\begin{table*}[t!]
\centering
\caption{Dataset Statistics.}
\label{tbl:dataset}
\begin{tabular}{lccc}
\hline
                  & Train                   & Validation              & Test                     \\ \hline
Review dates     & 1 Oct. 21 to 31 Jan. 22 & 1 Feb. 22 to 28 Feb. 22 & 1 Mar. 22 to 31 March 22 \\
Number of reviews & 1,176,261                 & 422,267                  & 570,588                   \\
Min. votes per review        & 0                       & 0                       & 0                        \\
Max. votes per review        & 44,178                   & 50,476                   & 47,798                    \\
Average votes per review     & 13.3                    & 13.65                   & 9.09                     \\
Number of Apps    & 9,572                    & 8,350                    & 8,369                     \\ \hline
\end{tabular}
\end{table*}

\section{Experimental Setup}
\label{sec:experiments}
In this section, we describe the dataset details along with data collection and pre-processing steps, evaluation metrics, and competing methods.
\subsection{Dataset}
\label{subsec:dataset}
\begin{figure}[t!]
\centering
\includegraphics[width=\linewidth]{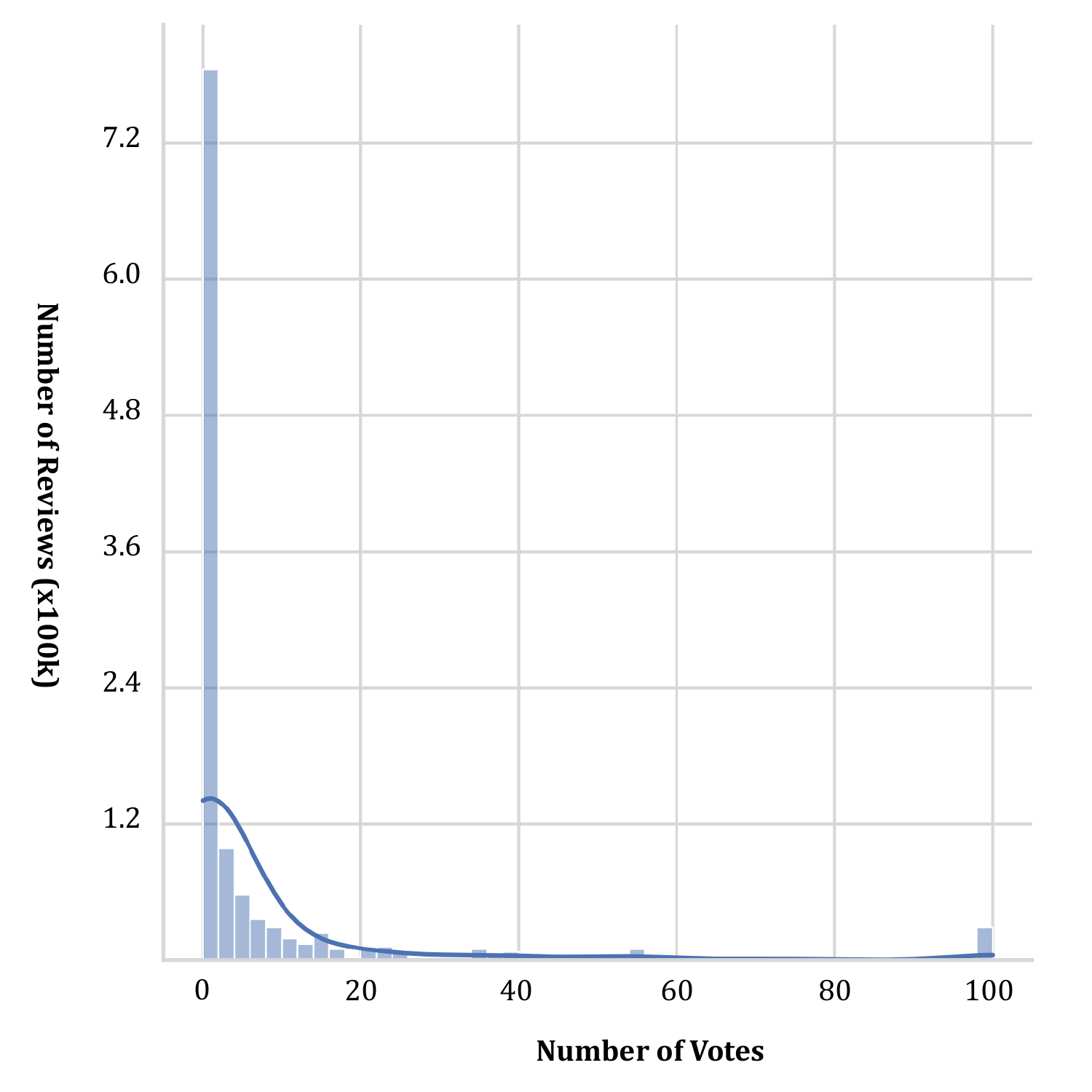}
\caption{Distribution of the review votes.}
\label{fig:distribution}
\vspace{-5pt}
\end{figure}

\stitle{Data Collection and Preprocessing.}
We collected over 2.1 million negative user reviews (i.e., reviews with ratings 1 and 2 only) from 9869 apps across 48 categories from Google Play. 
The reviews range from October 01, 2021, to March 31, 2022. Additionally, we tracked the number of votes each review received after one month, as the number of votes a user review receives is also determined by how long it has been since it was posted. 
For example, if a user review is posted on November 01, 2021, we record its number of votes on December 01, 2021.
The novelty of our new dataset is that many apps are not covered in existing user review datasets~\cite{rrgen,aarsyth}, and reviews are not recorded after a fixed timeframe.
Moreover, for each user review, we collected other useful features such as app name, app category, the total number of reviews for the app, price, and content rating, among others. 
Noise is often present in user reviews~\cite{gao2019emerging}. 
In order to minimize the noise in the dataset, we followed the best practices for data filtering~\cite{aarsyth}.
Table~\ref{tbl:dataset} presents important statistics about the dataset along with training, validation, and test split details and Figure~\ref{fig:distribution} shows the distribution for the number of votes (clipped to 100), and reviews received in a month. 
%It is important to highlight that the test set contains about 200 apps that were not part of the training set. 

\stitle{Dataset Analysis.}
Based on a manual analysis of a subset of the dataset (500 user reviews), we found that a small number of reviews received a large number of votes.
Additionally, we noticed that high-voted reviews often highlight specific app issues, and the concerns are actionable.
This highlights many users of an app face similar (or the same) app issues, and those issues should be prioritized following the maximum return criterion.
On the other hand, there is a large number of reviews that do not receive any votes. Moreover, such reviews do not contain any actionable complaints. Figure~\ref{fig:intro} presents a few sample user reviews.

\subsection{Evaluation}
\label{subsec:evalmetric}
We formulate the problem as binary classification as well as multi-class classification.
In binary classification, we consider whether a given user review will receive more than 100 votes in a month or not. Similarly, in the multi-class classification task, we consider five classes: 0 vote, 1-5 votes, 6-25 votes, 26-100 votes, and 100+ votes.
The effectiveness of {\ourmodel} is evaluated using both quantitative automatic metrics and subjective human studies.

\stitle{Automatic Metrics.}
Accuracy, F1 score, and MCC score are standard metrics for the quantitative evaluation of classification tasks. 
In our evaluation, we use macro F1 score which is a better metric for class imbalance tasks.
Moreover, we also use MCC score which has been proven to be a more robust and reliable metric.

\stitle{Human Study.}
To conduct a human study, we recruited 10 graduate students each with at least 4 years of app development experience. 
We randomly selected 100 user reviews weighted by their corresponding votes (clipped to 100) and assigned each review to three different students and asked them to classify the review whether it highlights any critical issue or not. 
% For a given review, if at least two (out of three) students agreed that the review highlights a critical issue, we assign it a label of 1, otherwise 0, and consider these labels as ground truth for human study.
% In this process, we ended up having 13 out of 100 user reviews with critical issues. 
% To compare the performance of {\ourmodel} with real software developers, we use binary classification task and see if {\ourmodel} predicts whether a given review will receive more than 100 votes, we consider it a critical issue.
% Similarly, we recruited another set of 10 graduate students with similar experience and asked them to predict whether a given user review is critical or not.

\subsection{Competing Approaches}
\label{subsec:competing}
We conduct extensive experimental evaluations using the large-scale dataset and compare it against several state-of-the-art transformer architectures~\cite{devlin2018bert,brown2020language,song2020mpnet,raffel2020exploring} that leverage several strategies to overcome the issue of class imbalance.  Similarly, we also compare against state-of-the-art social media popularity prediction models~\cite{saeed2022framework,khoerunnisa2022prediction}. In the following, we provide a brief overview of the competing models.

\begin{description}[leftmargin=1.2\parindent,labelindent=-5pt, itemsep=-0pt]
\item \textbf{BERT~\cite{devlin2018bert}}: BERT follows transformers architecture, uses only the encoder stack, and is considered a state-of-the-art model for many NLP tasks. We use several variants of the BERT for comparisons, such as the one that employs binary cross-entropy loss (BCE) and cross-entropy loss (CE). Moreover, we consider class imbalance strategies such as imbalanced dataset sampler, dice loss, and focal loss~\cite{lin2017focal,sudre2017generalised}. 
Finally, we also consider another variant of BERT where we get the last hidden layer's output and use it as a representation for the user review. We make predictions using weighted KNN that assigns weights to the neighbors by their distances. In our comparison, we use the pre-trained \myspecial{DistilBert} model.

\item \textbf{GPT-2~\cite{brown2020language}}: GPT-2 is based on the decoder stack of the transformers architecture. We use \myspecial{distilgpt2} in our comparison. For GPT-2 too, we use all the strategies, employed for BERT.

\item \textbf{T5~\cite{raffel2020exploring}}: The T5 model follows text-to-text architecture (both encoder and decoder stacks). In contrast to our proposed framework, this T5 baseline does not employ contrastive training and instead is trained using the same strategies as mentioned above for BERT and GPT-2. For comparison, we use the same model \myspecial{t5-base} that is employed in {\ourmodel} as the backbone model. This baseline also highlights the importance of contrastive training in our proposed framework.

\item \textbf{MPNet~\cite{song2020mpnet}}:
MPNet leverages the best of masked language modeling (e.g., BERT~\cite{devlin2018bert}) and permuted language modeling (e.g., XLNet~\cite{yang2019xlnet}) for pre-training. For comparison, we employ all the above-mentioned training strategies for MPNet as well. 

\item \textbf{EUSBoost~\cite{khoerunnisa2022prediction}}:
The state-of-the-art approach for predicting retweets employs evolutionary undersampling boosting strategy and uses several features including the user's network, the content of the tweet, and time, among others.

\item \textbf{DTP~\cite{saeed2022framework}}:
The state-of-the-art approach for predicting news popularity relies on initial tweeting behavior and temporal characteristics.

\end{description}

\begin{table}[t!]
\centering
\caption{Results of the binary classification task.}
\label{tbl:binary_results}
\begin{tabular}{llccc}
\hline
\textbf{Model}                  & \textbf{Approach}                         & \textbf{Accuracy} & \textbf{F1}      & \textbf{MCC}     \\ \hline
\multirow{5}{*}{BERT~\cite{devlin2018bert}}  & BCE Loss                & 0.5934   & 0.0612  & 0.1297  \\
                       & Imbalanced Sampler         & 0.7873   & 0.1059  & 0.1958  \\
                       & Dice Loss                        & 0.8719   & 0.1451  & 0.2252  \\
                       & Focal Loss                       & 0.9584   & 0.0958  & 0.0855  \\
                       & KNN                              & 0.5575   & 0.0592  & 0.1215  \\ \hline
\multirow{5}{*}{GPT-2~\cite{brown2020language}} & BCE Loss                & 0.6702   & 0.0981  & 0.1748  \\
                       & Imbalanced Sampler         & 0.8887   & 0.1750   & 0.2664  \\
                       & Dice Loss                        & 0.9099   & 0.2230   & 0.2773  \\
                       & Focal Loss                       & 0.8960    & 0.1033  & 0.0914  \\
                       & KNN                              & 0.5125   & 0.1125  & 0.2698  \\ \hline
\multirow{5}{*}{T5~\cite{raffel2020exploring}}    & BCE Loss                & 0.4705   & 0.0393  & 0.1026  \\
                       & Imbalanced Sampler         & 0.9405   & 0.2474  & 0.2443  \\
                       & Dice Loss                        & 0.8078   & 0.1347  & 0.2127  \\
                       & Focal Loss                       & 0.9394   & 0.0876  & 0.1022  \\
                       & KNN                              & 0.5642   & 0.0580   & 0.1295  \\ \hline
\multirow{5}{*}{MPNet~\cite{song2020mpnet}} & BCE Loss                & 0.5276   & 0.0745  & 0.1205  \\
                       & Imbalanced Sampler         & 0.7163   & 0.1372  & 0.1887  \\
                       & Dice Loss                        & 0.8217   & 0.1820   & 0.2525  \\
                       & Focal Loss                       & 0.9511   & 0.1016  & 0.1166  \\
                       & KNN                              & 0.5298   & 0.1270   & 0.2446  \\ \hline
EUSBoost~\cite{khoerunnisa2022prediction}                  & Undersampling    & 0.2664  & 0.0355 & 0.0661  \\
DTP~\cite{saeed2022framework}                  & Temporal Propagation     & 0.2430  & 0.0395 & 0.0690 \\ \hline
\multirow{2}{*}{\ourmodel}  & KNN     & \textbf{0.9798}   & 0.3201  & 0.3108  \\
                       & RNC        & 0.9784   & \textbf{0.3465}  & \textbf{0.3409}  \\ \hline
\end{tabular}
\end{table}

\section{Results}
\label{sec:evals}
In this section, we discuss the quantitative results as well as the findings of the human study. Moreover, we also present the results of the ablation study.

\stitle{Automatic Metrics.}
Table~\ref{tbl:binary_results} presents the results of the binary classification task, where we consider whether a given user review will receive more than 100 votes or not. 
The results that perform the best for each metric have been highlighted in bold. 
We see that {\ourmodel} with weighted KNN inference performs the best for the accuracy metric.
Whereas {\ourmodel} with radius neighbor classification strategy performs the best for F1 score as well as MCC score. Both MCC score and macro-averaged F1 score are considered more robust metrics for class imbalance tasks.
It is important to highlight that the best approach among the competitors for the F1 score is T5 with an imbalanced dataset sampler, which is roughly 10 percentage points below {\ourmodel} (0.2474 vs 0.3465).
Similarly, GPT-2 with dice loss performs the best among the competitor approaches for MCC score that is over 6 percentage points lower than our proposed framework. 
This performance gain over state-of-the-art models that follow transformers architecture can be attributed to the training strategies such as self-supervised and contrastive training.
The self-supervised training in phase one of {\ourmodel} adapts the model to the new distribution of the dataset (i.e., user reviews) in an unsupervised way.
Similarly, phase two performs contrastive training that proves to be robust against the class imbalance challenge, and produces generic and accurate representations of the user reviews.  
Moreover, we also noticed that the contrastive training strategy is robust against the continuous distributional shift among the issues in the apps since there are over 200 apps in the test set that were not part of the training set.
Moreover, we also pay attention to the worst performance of state-of-the-art models for social media posts popularity prediction ESUBoost and DTP.
The lack of features such as the user's network and temporal interactions could be the cause of their worst performance.

\begin{table}[t!]
\centering
\caption{Results of the multi-class classification task.}
\label{tbl:multiclass_results}
\begin{tabular}{llccc}
\hline
\textbf{Model}                  & \textbf{Approach}                 & \textbf{Accuracy} & \textbf{F1}      & \textbf{MCC}     \\ \hline
\multirow{5}{*}{BERT~\cite{devlin2018bert}}                   & CE Loss        & 0.5377   & 0.2914  & 0.2355  \\
                       & Imbalanced Sampler & 0.5543   & 0.3004  & 0.2651  \\
                       & Dice Loss                & 0.5660    & 0.3425  & 0.2392  \\
                       & Focal Loss               & 0.5618   & 0.3340   & 0.1212  \\
                       & KNN                      & 0.5062   & 0.2989  & 0.2232  \\ \hline
\multirow{5}{*}{GPT-2~\cite{brown2020language}} & CE Loss        & 0.4566   & 0.1985  & 0.1938  \\
                       & Imbalanced Sampler & 0.6109   & 0.3512  & 0.2975  \\
                       & Dice Loss                & 0.5710    & 0.3438  & 0.2514  \\
                       & Focal Loss               & 0.5557   & 0.3115  & 0.1397  \\
                       & KNN                      & 0.5023   & 0.2886  & 0.2775  \\ \hline
\multirow{5}{*}{T5~\cite{raffel2020exploring}}    & CE Loss        & 0.4257   & 0.1980   & 0.1792  \\
                       & Imbalanced Sampler & 0.5749   & 0.3549  & 0.2810   \\
                       & Dice Loss                & 0.5248   & 0.3161  & 0.2309  \\
                       & Focal Loss               & 0.5503   & 0.2963  & 0.1489  \\
                       & KNN                      & 0.5094   & 0.3029  & 0.2459  \\ \hline
\multirow{5}{*}{MPNet~\cite{song2020mpnet}} & CE Loss        & 0.3591   & 0.1521  & 0.1357  \\
                       & Imbalanced Sampler & 0.4899   & 0.2726  & 0.2058  \\
                       & Dice Loss                & 0.5152   & 0.2865  & 0.2285  \\
                       & Focal Loss               & 0.5928   & 0.3051  & 0.1795  \\
                       & KNN                      & 0.5167   & 0.3311  & 0.2507  \\ \hline
EUSB~\cite{khoerunnisa2022prediction}                  & Undersampling          & 0.3121  & 0.1265  & 0.0913 \\
DTP~\cite{saeed2022framework}                  & Temporal Propagation      & 0.3189  & 0.1115 & 0.1045 \\ \hline
\multirow{2}{*}{\ourmodel}  & KNN & \textbf{0.6687}   & 0.4211  & 0.3442  \\
                       & RNC    & 0.6685   & \textbf{0.4245}  & \textbf{0.3455}  \\ \hline
\end{tabular}
\end{table}

Table~\ref{tbl:multiclass_results} shows the results of the multi-class classification task, where we formulate the task as a 5-class classification task.
The main reason behind considering this task is that oftentimes developers might want to consider how critical a given user review will become on a scale of 1 to 5, instead of a binary prediction.
On the multi-class classification task, our proposed {\ourmodel} is a clear winner on all three quantitative metrics, accuracy, F1 score, and MCC score. 
Specifically, {\ourmodel} with KNN performs the best on the accuracy metric that is over five percentage points better than the best competitor (i.e., GPT-2 with an imbalanced dataset sampler).
Similarly, on the F1 score, {\ourmodel} with RNC outperforms the best competing approach (T5 with imbalanced data sampler) by 19.61\%.
If we consider the MCC score, GPT-2 with an imbalanced data sampler is the best competing approach that is about five percentage points worse than {\ourmodel} with RNC.
We also notice that among the competing approaches there is no clear winner. 
However, an imbalanced dataset sampling strategy that seems a very simple approach tends to work better for the most part.
Our contrastive training strategy employed in phase two mimics imbalance dataset sampling in a way that exposes minority class instances more often to the model by negative sampling, though contrastive learning has other advantages as well.
Approaches like EUSB and DTP are once again among the worst performing models for the multi-class classification task because of the same reasons (as binary classification task).
We also investigated the effect of the dice and focal loss functions that have been specifically designed to overcome the issue of class imbalance in the datasets. 
We noted that these loss functions are almost always better than cross-entropy or binary cross-entropy loss functions which validates that these loss functions take care of the class imbalance issue to some extent.
To summarize the results of the quantitative evaluation, {\ourmodel} is a clear winner on both tasks on all the metrics.

The readers might not be impressed with the accuracy of {\ourmodel} which is slightly over 66\% for the multi-class classification task. 
We also dig deeper into this and found out that most of the mistakes are on the boundary of classes upon further error analysis. 
For example, reviews that receive 25 votes and 26 votes, respectively, are not any different from each other, but they belong to different classes in our dataset based on the class partition boundary. 
Moreover, the value of $\lambda$ is set to four in our experiments, which does not contrast the reviews with the number of votes difference less than four. 
To overcome such cases, we also tried smaller values of $\lambda$ that addressed this issue, but overall performance deteriorated.
Finally, we conducted another analysis where we considered whether the top two predicted classes are among the true class or not. In this experiment, {\ourmodel} was able to achieve an accuracy of over 98\%.

\stitle{Ablation study.}
We specifically employed T5 baseline model as a competitor to highlight the improvement of our proposed training strategies.
We notice that there is always a significant performance difference between {\ourmodel} and all the variants of the T5 baseline on both binary as well as multi-class classification tasks for all metrics.
Moreover, we also introduced a KNN-based variant for all the transformers models to point out that the performance gain in {\ourmodel} is not due to off-the-shelf implementation of KNN.
Specifically, we used the last hidden layer's output of all the models and further used KNN for inference. 
Experiments based on T5 and incorporation of KNN component in all the competing models highlight that {\ourmodel}'s performance gain is mainly due to the contrastive loss and self-supervised training.

\stitle{Human Study.}
For human study, for a given review, when at least two (out of three) students agreed that the review highlights a critical issue, we assign it a label of 1, otherwise 0, and consider these labels as ground truth.
In this process, we ended up having 13 out of 100 user reviews with critical issues. 
To compare the performance of {\ourmodel} with real software developers, we used a binary classification task and see if {\ourmodel} predicts whether a given review will receive more than 100 votes. If so, we consider it a critical issue.
Similarly, we recruited another set of 10 graduate students with similar experience and asked them to predict whether a given user review is critical or not.
Based on the experiment of the human study, we found out that {\ourmodel} achieves competitive performance to the experienced app developers, since {\ourmodel} made only three mistakes (out of 100 predictions) in predicting critical app issues as compared to two mistakes by humans.
\section{Related Work}
\label{sec:related}

\stitle{Software Feature Prioritization.}
Recent works studied software release planning including feature improvements and fixing issues by utilizing user feedback. 
Maalej et al.~\cite{maalej2015toward} discussed state-of-the-art and future directions for data-driven requirement engineering.
Ciurumelea and others~\cite{ciurumelea2017analyzing} developed a classification system for app reviews, furthermore to resolve issues mentioned in reviews they recommend a recommendation system to suggest code changes. % 
Nayebi and Ruhe~\cite{nayebi2017optimized} utilize software feature properties such as value and cohesiveness to select features for the next release planning. 
Later, they proposed Asymmetric Release Planning (ARP)~\cite{nayebi2018asymmetric}, where they model asymmetric feature evaluation and formulate release planning as a bi-creation optimization planning.
More recently, Zhang and others~\cite{zhang2019software} studied user reviews and formulate software feature prioritization as an
optimization problem that targets maximizing app ratings for a certain user group.
Yang et al.~\cite{yang2021phrase} use a semi-automated approach to prioritize bugs to fix, they study 6 apps and extract fourteen features to classify reviews. 

In comparison, our approach develops on a much larger dataset as compared to existing studies and develops a novel way to utilize user likes and dislikes which has been ignored by existing work. Furthermore, our work complements many existing techniques, e.g., the combination of code changes recommendation in~\cite{ciurumelea2017analyzing} and {\oursys} could gain more benefits.

\stitle{Social Posts Popularity Prediction.} 
Many studies focus on social media platforms to predict the popularity of posts and analyze social behavior~\cite{lee2015will}. 
Nesi et al.~\cite{nesi2018assessing} developed a predictive model to anticipate whether a certain tweet will be retweeted by a user or not~\cite{nesi2018assessing}. 
Tsugawa and Kito\cite{tsugawa2017retweets} utilize user retweets to find user relationships based using a prediction model.
Zhang et al.~\cite{zhang2011predicting} collected Twitter feed for six months to predict correlation to stock market performance including Dow Jones, NASDAQ, and S\&P 500.  
Daga et al.~\cite{daga2020prediction} predicted likes and retweets for Twitter posts. Similarly, Khoerunnisa et al.~\cite{khoerunnisa2022prediction} utilized the user profile, content, and time to predict retweets. Saeed and others~\cite{saeed2022framework} predicted the news popularity based on initial tweeting behavior on Twitter. 
On the contrary, {\oursys} focuses on user reviews on App stores, which do not contain user profiles compared to social networks such as Facebook and Twitter and this work bridges this gap. Furthermore, we predict the likability of user reviews which is not explored by existing work.

\stitle{Analysis of App Reviews.} 
Many existing studies analyze app reviews from different perspectives, including app feature requests~\cite{gu2015parts}, app store analysis~\cite{martin2016survey}, developer response generations~\cite{rrgen,aarsyth} and so on.
Martin and others~\cite{martin2016survey} studied app stores, they concluded that app reviews are important and contain useful information. 
\textsc{Mara}~\cite{iacob2013retrieving} utilized app reviews to predict feature requests by the users. 
Gu and Kim~\cite{gu2015parts} developed an app review parser to find out user opinions on different app features.
A study~\cite{hassan2018studying} on $4.5$ million reviews observed the importance of developer replies to app reviews. 
Similarly, a Google study~\cite{googleplayresponse} highlighted that developer responses to reviews result in better app ratings. {\oursys} extends the findings of existing research by providing a large dataset of reviews and prioritizes app issues based on user reactions. 
\balance
\section{Conclusion}
\label{sec:conclusion}
We have presented {\oursys}, a novel framework to automatically prioritize app issues to allow developers to proactively resolve critical app issues. 
{\oursys} is motivated by the impossibility to process a large and continuously growing number of user reviews at a rapid pace. 
It enables app developers to automatically identify critical app issues that are important to resolve for users' satisfaction, thanks to the accurate prediction power of the proposed {\oursys}. 
Our proposed framework relies on the power of pre-trained language models and self-supervised learning strategies. 
The contrastive training in phase two of the framework produces generic, accurate, and robust representations of the user reviews and empowers simple approaches like KNN to make reliable predictions. Moreover, we leverage the FAISS library to speed up the inference.
To conduct extensive experimental analysis, we crawled a large-scale dataset of over 2.1 million negative user reviews from about 10,000 Google Play apps.
Our extensive experimental evaluations demonstrate that {\ourmodel} outperforms a wide range of transformers architecture such as BERT, GPT, T5, and MPNet on well-accepted quantitative metrics consistently.
Moreover, our approach proved to be better than state-of-the-art social media posts' popularity prediction approaches by a large margin. 
Specifically, {\ourmodel} is at least 27.97\% and 24.50\% better than all the competitors on the MCC score.
Last but not least, our human study shows that the performance of {\ourmodel} is as good as senior app developers.

%\clearpage
\bibliographystyle{plain}
\bibliography{sample-base.bib}

\end{document}